\documentclass[12pt,a4paper]{article}
\pdfoutput=1
\usepackage[utf8]{inputenc}
\usepackage[ngerman,english]{babel}
\usepackage{color}
\usepackage{listings}
\lstdefinelanguage{Sage}[]{Python}{morekeywords={as,False,sage,True,append,bits,BooleanPolynomialRing,extend,flatten,gens,groebner_basis,ideal,inject_variables,join,latex,maximal_degree,monomials,nmonomials,PolynomialRing,quotient_ring,split,subs,substitute,update,Sequence,ZZ},sensitive=true}
\definecolor{dblackcolor}{rgb}{0.0,0.0,0.0}
\definecolor{dbluecolor}{rgb}{0.01,0.02,0.7}
\definecolor{dgreencolor}{rgb}{0.2,0.4,0.0}
\definecolor{dgraycolor}{rgb}{0.30,0.3,0.30}

\lstnewenvironment{sageverbatim}[1][numbers=none]{\lstset{language=Sage,#1}}{}
\usepackage{graphicx}
\usepackage{verbatim}
\usepackage{multirow}
\usepackage[hyphens]{url}
\usepackage[T1]{fontenc}
\usepackage{lmodern}
\usepackage[bookmarks,bookmarksnumbered,colorlinks=true]{hyperref}
\newcommand{\mod}{\mathop{\rm mod}\nolimits}
\newcommand{\Byt}[1]{{\mathbf{#1}}}
\newcommand{\hexa}[1]{{\textsf{\small '#1'}\,}}
\newcommand{\etal}{\emph{et al.}}
\title{Generating and Exploring S-Box Multivariate Quadratic Equation Systems with SageMath}
\author{A.-M. Leventi-Peetz$^1$ and J.-V. Peetz$^2$
  \\*[1ex]
  \parbox{20em}{\center$^1$\,\small
    Federal Office for Information Security,\\
    Godesberger Allee 185--189, DE-53175 Bonn, Germany\\
    \url{leventi-peetz@bsi.bund.de}}
  \\*[1ex]
  \parbox{20em}{\center$^2$\,\small
    DE-53343 Wachtberg, Germany\\
    \url{jvpeetz@web.de}}
}
\date{May, 2017}
\hypersetup{
  pdfinfo={
    Author={A.-M. Leventi-Peetz, J.-V. Peetz},
    Title={Generating and Exploring S-Box Multivariate Quadratic Equation Systems with SageMath}
  }
}
\begin{document}
\maketitle
\begin{abstract}
A new method to derive Multivariate Quadratic equation systems (MQ) for the input and output bit variables of a cryptographic S-box from its algebraic expressions with the aid of the computer mathematics software system SageMath is presented.  We consolidate the deficiency of previously presented MQ metrics, supposed to quantify the resistance of S-boxes against algebraic attacks.
\\*[1ex]
\noindent\textbf{Key Words} --
Algebraic attack resistance, algebraic cryptanalysis, Lagrange polynomial, multivariate quadratic polynomial equation system, polynomial quotient ring, SageMath, SAT solver, S-box, Rijndael AES.
\end{abstract}
\section{Overview}
We present a new automated way to produce and investigate Multivariate
Quadratic equation systems (MQ) over $GF(2)$ for the
Rijndael S-box (S$_{\mathrm{RD}}$) and alikes.
In the next we first shortly survey the principles of S$_{\mathrm{RD}}$
in section \ref{s:SboxRD}.
Recently, Jie Cui \etal~\cite{cui2014:aes} claimed to have presented
a new and concise approach for generating such MQ for S$_{\mathrm{RD}}$
and also to have proposed a cryptographically more secure S-box.
In section \ref{s:aboutCui} we depict the derivation of Cui et al.
In section \ref{s:SboxSage} we present our automated way to produce
the MQ for S$_{\mathrm{RD}}$ aided by the computer mathematics
software system SageMath~\cite{stein2015:sage}.
As we demonstrate later, this method can be applied for other
proposed S-boxes, alleged to be cryptographically stronger.
We also generate the Gröbner bases describing the two different
S-boxes and calculate solutions of all equation systems with the help
of a SAT solver.
In section \ref{s:Algebraicattacks} we investigate, critically
discuss, and dismiss equation-metrics-based criteria as inappropriate
to estimate the resistance against algebraic attacks (RAA) of an S-box.
In sections \ref{s:MQofAIA} and \ref{s:MQforSAIA} we build MQ for the
S-box proposed by Jie Cui \etal~\cite{cui2014:aes} and show why
this is not an improvement.
The complicated algebraic expression for the S-box constructed by
Cui~\etal\ leads to a great number of equations and independent
monomials in the resulting MQ.  On the basis of the RAA formulas the
new S-box should demonstrate a remarkable robustness against algebraic
attacks.
But from that same S-box can be derived a much simpler MQ leading to a
polynomial system nearly as easy solvable as that of the original
S$_{\mathrm{RD}}$. This is what we do in section \ref{s:MQforSAIA}
greatly facilitated by SageMath.

Our main contribution is the demonstration of the handiness which
Sagemath offers to the researcher so that he can derive his MQ and its
metrics in a fast way and transparently verify the quality of existing
formulas supposed to quantify resistance of the MQ to algebraic
attacks.
In conclusion, we couldn't validate the predicted reduced hardness for
a system to solve with increasing number of equations and number of
independent monomials in the polynomial systems according to suggested
RAA formulas.
\section{Principle of Rijndael S-box RD}
\label{s:SboxRD}
We shortly repeat its well known principles and algebraic
properties~\cite{courtois2002:algres}.
Looking upon 8-bit bytes as elements in $GF(2^8)$, Rijndael's S-box is
a mapping  $S:
GF(2^8)\longrightarrow GF(2^8)$ in form of a combination of an inverse
function $I(\Byt{x})$ which is a multivariate inverse modulo the
irreducible polynomial $m(t)=t^8+t^4+t^3+t+1$ and an affine
transformation function $A(\Byt{x})$. $\Byt{x}$ is a byte variable
consisting of bits $x_i(i = 0, \ldots , 7)$, with $x_7$ symbolizing the
most significant bit: $\Byt{x} = \sum_{i=0}^{7} x_i t^i$.
The modular inverse function $I(\Byt{x})$ is defined as:
\begin{equation}
  I(\Byt{x}) = \Byt{x}^{254} \mod m(t)
\end{equation}
i.e., the modular inverse of 0 is mapped to 0.
According to the AES design~\cite{DR2002:DRA,FIPS2001:AES},
the affine transformation $A(\Byt{x})$ can also be described as a
modular polynomial multiplication followed by an addition (XOR) of a
constant polynomial:
\begin{equation}
  A(\Byt{x}) = \Byt{a}\, \Byt{x} \mod (t^8 + 1) + \Byt{b}
  \label{e:affx}
\end{equation}
with $\Byt{a}=\hexa{1F}$ and $\Byt{b}=\hexa{63}$.
A two-digit hexadecimal number stands for a \emph{constant} byte,
that is a polynomial in $t$, e.g, $\hexa{63}$ for $t^6 + t^5 + t + 1$.
The Rijndael S-box can be written as:
\begin{equation}
  S_{\mathrm{RD}}(\Byt{x}) = A \circ I = A(I(\Byt{x}))
  \label{e:sboxcomp}
\end{equation}
\section{Rijndael S-box explored by Cui et al.}
\label{s:aboutCui}
Cui \etal~\cite{cui2014:aes} (as Courtois and
Pieprzyk~\cite{courtois2002:algres} before them) utilize the
Rijndael S-box composition (\ref{e:sboxcomp}) to derive an MQ for it.
With $\Byt{x}$ the input and $\Byt{z}$ the output value, and the
intermediate variable $\Byt{y} = I(\Byt{x})$ they note:
$\Byt{z} = S_{\mathrm{RD}}(\Byt{x}) = A(\Byt{y}) = A(I(\Byt{x}))$.
Considering the inverse transformation $\Byt{y}=I(\Byt{x})$,
obviously $\Byt{x}\Byt{y} = 1$
when $\Byt{x}$ not equal $0$, which reads in polynomial form:
\begin{equation}
  \left(\sum_{i=0}^{7} x_{i} t^{i}\right) \left(\sum_{j=0}^{7} y_{j} t^{j}\right)
  \mod m(t) = 1
\end{equation}

The above modulo division is then analytically performed and a
comparison of coefficients of terms of the same order
in $t^k$, $0 \le k \le 7$,
leads to the first eight multivariate quadratic equations for Rijndael
S-box on the pages 2483, 2484 of the paper of
Cui \etal~\cite{cui2014:aes}
The authors give all the steps and in-between
results of the complete length of the calculation. They formulate and
evaluate two additional equations of the byte variables to define
the S-box completely. Doing so, Cui \etal\ replicate results already
presented in 2002 by Courtois and Pieprzyk in the extended version
of~\cite{courtois2002:algres}.
\section{Rijndael S-box coded in Sage}
\label{s:SboxSage}
%
%
SageMath or briefly Sage (System for Algebra and Geometry
Experimentation) is a free open-source software system for computer
mathematics~\cite{wikipedia:SageMath}. It is licensed under
the Gnu General Public License. It builds on top of many existing
computer mathematics open-source packages. Their combined power is
accessible through a common, Python-based language interface from the
command line or a web browser.
Originally, it is designed by William Stein, still the leader of the
SageMath project, and also inventor of
SageMathCloud~\cite{SageMathCloud} for collaborative computational
mathematics.

In order to work with polynomials like $\Byt{x} = \sum_{i=0}^{7} x_{i} t^{i}$,
Sage provides modules to construct rings of multivariate polynomials.
The polynomials $\Byt{x}$, $\Byt{y}$, and $\Byt{z}$ introduced in the
previous section we model in Sage as follows.\footnote{The complete
  code presented here together with its output is
  accessible at SageMathCloud~\cite{SBoxinSMC}.} 
In line 1 of listing~\ref{c:polyring} a variable for the number of
bits is defined for convenience.
In line 2 a list of strings for the names of the coefficients of the three
byte polynomials is generated (\lstinline|['x0', 'x1', .., 'z7']|).

\begin{sageverbatim}%
[caption={Byte polynomials over a quotient ring},label=c:polyring]
  nb = 8
  varl = [c + str(p) for c in 'xyz' for p in range(nb)]
  B = BooleanPolynomialRing(names = varl)
  B.inject_variables()
  P.<p> = PolynomialRing(B)
  Byte.<t> = P.quotient_ring(p^8 + p^4 + p^3 + p + 1)
  X = B.gens()[:nb]
  Y = B.gens()[nb:2*nb]
  x = sum([X[j]*t^j for j in range(nb)])
  y = Byte(list(Y))
\end{sageverbatim}
In line 3 a Boolean polynomial ring for these coefficients is
constructed which assigns $GF(2)$ properties to them.
In line 4 the coefficient names are made available as
variables.
In line 5 a polynomial ring over the Boolean polynomial ring
\lstinline|B| is constructed and from that, in line 6, the final
quotient ring \lstinline|Byte| with modulus $m(t)$.
In lines 7 and 8, lists\footnote{To be exact, in Python these are
  \emph{tuples}, i.e., immutable \emph{lists}.}
of coefficient variables of the byte
polynomials are created for convenience.
With the help of these lists, in the last two lines the
polynomials are modeled in two equivalent ways, $\Byt{x}$ explicitly,
and $\Byt{y}$ by using the \lstinline|Byte| constructor.

Now one can already evaluate the product $\Byt{x}\Byt{y}$ in
Sage with the commands:
\begin{sageverbatim}
  E3 = x * y
  eqs3 = E3.list()
\end{sageverbatim}
In the second line we used the \lstinline|list()| attribute to get the
coefficients of each power of $t$ in expression \lstinline|E3|.
Due to the usage of the quotient ring, \lstinline|E3| is of degree 7, the length of list \lstinline|eqs3| (the number of coefficients) is 8. The terms we have gotten with Sage compare with the right-hand sides (rhs) of the system of equations with number (3) in the paper of Cui \etal~\cite{cui2014:aes}


Cui \etal\ proceeded with the generation of the next set of
equations for Rijndael's S-box, the affine transformation.
From equation (\ref{e:affx}) setting $\Byt{z} = A(\Byt{y})$ it follows
\begin{equation}
  \Byt{y} = \Byt{a}^{7} (\Byt{z} + \Byt{b}) \mod (t^8 + 1)
  \label{e:invaffx}
\end{equation}
since $\Byt{a}^{7}\, \Byt{a} \mod (t^8 + 1) = 1$.
Substituting (\ref{e:invaffx}) for $\Byt{y}$ in $\Byt{x}\Byt{y}$
we get the final form of the first implicit eight equations
representing Rijndael S-box.
In Sage the values substitution is accomplished with the help of a
so called \emph{dictionary}.
By using equation (\ref{e:invaffx}) the code in
Listing~\ref{c:invaffx} generates this dictionary,
called \lstinline|eqs4|.
\begin{sageverbatim}[numbers=none,caption={Generate a dictionary to substitute $\Byt{y}$ variables},label=c:invaffx]
  Baff.<u> = P.quotient_ring(p^8 + 1)
  Z = B.gens()[2*nb:][:nb]
  z = Baff(list(Z))
  a = u^4 + u^3 + u^2 + u + 1
  b = u^6 + u^5 + u + 1
  eqs4 = dict(zip(Y, (a^7 * (z + b)).list()))
\end{sageverbatim}
The first line sets up a quotient ring modulo $t^8 + 1$. The next four
lines define the byte variable $\Byt{z}$ and the two constant
polynomials $\Byt{a}$ and $\Byt{b}$ in this ring
(with generator \lstinline|u|).
The rhs of equation (\ref{e:invaffx}) simply reads
\lstinline|a^7 * (z + b)| in the code. The dictionary is constructed
in the last line. The result is shown in Listing~\ref{c:ydict}.
\begin{sageverbatim}[numbers=none,caption={Dictionary to substitute $\Byt{y}$ variables},label=c:ydict]
  {y7: z6 + z4 + z1,
   y6: z5 + z3 + z0,
   y5: z7 + z4 + z2,
   y4: z6 + z3 + z1,
   y3: z5 + z2 + z0,
   y2: z7 + z4 + z1 + 1,
   y1: z6 + z3 + z0,
   y0: z7 + z5 + z2 + 1}
\end{sageverbatim}

The substitution of $\Byt{y}$ in equation \lstinline|eqs3| via the
dictionary succeeds with the following:
\begin{sageverbatim}
  eqs5 = [_.subs(eqs4) for _ in eqs3]
\end{sageverbatim}
The result is again a list, the members of which give the first set of
eight multivariate quadratic equations of the S-box by setting the
byte variable product equal to 1.
This list of terms \lstinline|eqs5| is identical to the system
of equations (5) of Cui \etal\@ Those equations with zero constant term (7 out of 8
above) are true with probability equal to 1. The 8th equation (the coefficient of $t^0$) is true
only when $\Byt{x}\neq 0$, so that this equation is true with a
probability 255/256. Furthermore for $\forall\, \Byt{x} \neq 0$
$\Byt{x}=\Byt{x}^{2}\Byt{y}$.
Obviously this last equation is true also when $\Byt{x}=0$, so that one can write:
\begin{equation}
  \forall\, \Byt{x} \in GF(2^{8})
  \left\{\begin{array}{lcl}
      \Byt{x}    & = & \Byt{y} \Byt{x}^{2}\\
      \Byt{x}^{2} & = & \Byt{y}^{2} \Byt{x}^{4}\\
      & \vdots &\\
      \Byt{x}^{128} & = & \Byt{y}^{128} \Byt{x}^{256} = \Byt{y}^{128} \Byt{x}
      \end{array}
  \right.
\end{equation}
Cui \etal\ take the last of the above and write two symmetrical equations
to generate an additional set of 16 equations for the Rijndael S-box.
The equations they take are:
\begin{equation}
  \left\{ \
  \begin{array}{lcl}
    \Byt{x}^{128} & = & \Byt{y}^{128} \Byt{x}\\
    \Byt{y}^{128} & = & \Byt{x}^{128} \Byt{y}\\
  \end{array}
  \right.
  \label{e:xy128}
\end{equation}
We develop the two last equations to get the needed additional 16
equations for the implicated variables. We also substitute in these
equations $\Byt{y}$ by using the
dictionary in listing~\ref{c:ydict}.
\begin{sageverbatim}
  E7 = x^128 + y^128 * x
  eqs7 = [_.subs(eqs4) for _ in E7.list()]
\end{sageverbatim}
The result is a list of terms which are
practically the equations (7) of Cui \etal\ written in the reverse
order than that of Cui's paper. Cui \etal\ begin with the expression
corresponding to the highest order term of $E7$ while the sage list
begins with the constant term.
%
%
These terms are identical with the rhs of equations (7) in Cui
\etal~\cite{cui2014:aes} with a
couple of minimal differences which we attribute to typographical
errors in the paper of Cui \etal

Similarly, we write:
\begin{sageverbatim}
  E8 = y^128 + x^128 * y
  eqs8 = [_.subs(eqs4) for _ in E8.list()]
\end{sageverbatim}
and, by setting these terms equal to 0, get the next block of eight equations for the Rijndael S-box which are to be compared with the system (8) of Cui \etal\@
%
%
Here we see a couple of discrepancies which we again attribute to
typographical mistakes in the reference paper.

Using the Sage model of this S-box it is easy to count the number of
equations, the number of terms in each equation and determining the
minimal and maximal number of terms, as well as the total number of
different terms, as shown in Listing~\ref{c:surveySRD}.
\begin{sageverbatim}%
[numbers=none,caption={Survey of first S-box equation system},label=c:surveySRD]
  mq1 = eqs5[1:] + eqs7 + eqs8
  len(mq1)
  lmon1 = [len(_.monomials()) for _ in mq1]
  min(lmon1)
  max(lmon1)
  Sequence(mq1).nmonomials()
\end{sageverbatim}
As mentioned above, the first equation is discarded since it is only
true with probability 255/256 (false if $\Byt{x} = 0$).
This gives for the Rijndael S-box 23 equations, with between 28 and 49
monomials per equation and, in total, 81 different monomials.

Finding the 256 solutions of this equation system representing the
value table of the byte S-box with a SAT solver in
Sage is accomplished with the following two lines of code:
\begin{sageverbatim}%
[numbers=none,caption={SAT solver usage},label=c:SATsolver]
  from sage.sat.boolean_polynomials \
       import solve as solve_sat
  %time r = solve_sat(mq1, n=infinity)
\end{sageverbatim}
This takes ca.\ 0.6~s CPU time on a decent computer
(2.8 GHz CPU, 8 GB RAM).
As a point of reference we also evaluate the Gröbner basis (GB) of
this MQ and print the number of the basis equations, as well as the
maximal degree and the number of its monomials:
\begin{sageverbatim}%
[numbers=none,caption={Gröbner basis evaluation},label=c:GroebnerBasis]
  Idl1 = B.ideal(mq1)
  %time mq1gb = Idl1.groebner_basis()
  print len(mq1gb)
  print mq1gb.maximal_degree()
  print mq1gb.nmonomials()
\end{sageverbatim}
The evaluation of the Gröbner basis takes ca.\ 14 seconds, it has 8
equations of degree 7 with 263 different monomials.
The solution of the basis equations with the SAT solver is about 4
times as fast as the solution of the MQ.

Courtois and Pieprzyk~\cite{courtois2002:algres} state that these 23
equations are linearly independent.  Nonetheless, the last 16
equations (\ref{e:xy128}) only, i.e. \lstinline|mq2 = eqs7 + eqs8|,
are already sufficient to evaluate the Gröbner basis and to compute
the S-box value table of 256 solutions with the SAT solver which takes
ca.\ 0.7~s CPU time on the same computer. The 16 equations describing
the Rijndael S-box have between 28 and 49 monomials per equation and,
in total, 81 different monomials.
\section{Algebraic attacks and S-box optimization}
\label{s:Algebraicattacks}
For quantifying the resistance against algebraic attacks for $r$
equations in $t$ terms over $GF(2^{n})$ Cui \etal~\cite{cui2014:aes}
have used the criterion of
Cheon and Lee~\cite{cheon2004:algres} which defines the
Resistance against Algebraic Attacks (RAA) $\Gamma$ as:
\begin{equation}
  \Gamma = \left( \frac{t - r}{n} \right)^{\lceil (t-r)/n \rceil}
  \label{e:gamma}
\end{equation}
%
%
Courtois and Pieprzyk~\cite{courtois2002:algres} use another criterion
\begin{equation}
  \Gamma_{\mathrm{CP}} = \left( \frac{t}{n} \right)^{\lceil t/r \rceil}
  \label{e:gammaCP}
\end{equation}
%
(note the brackets $\lceil\, \rceil$ in the exponent indicating the
ceiling function).\footnote{Parameter $n$ could be interpreted as
  number of dependent variables, see section~\ref{s:MQofAIA}.}
The value of these criteria should reflect the difficulty of solving
multivariate equations.
For the Rijndael S-box we counted 23 equations and, in total, 81 different
monomials.
Therefore, it has $\Gamma = (29/4)^{8} \approx 2^{22.9}$ and
$\Gamma_{\mathrm{CP}} = (81/8)^{4} \approx 2^{13.4}$.
The 16 equations system has $\Gamma = (65/8)^{9} \approx 2^{27.2}$ and
$\Gamma_{\mathrm{CP}} = (81/8)^{6} \approx 2^{20.0}$. Compared with the
relation of the computational effort of the SAT solver for the MQ for
23 and 16 equations respectively, the $\Gamma$ values for the 16
equation system are exaggerated.

To generate a \emph{harder} to solve equation system Cui
\etal~\cite{cui2014:aes} have introduced a more complicated Rijndael
S-box structure which they name Affine-Inverse-Affine (AIA)
structure. This S-box will be explored in detail in the next two sections.
\section{MQ of the AIA structure S-box in Sage}
\label{s:MQofAIA}
In Cui, Huang, \etal\ (2011)~\cite{cui2011:AIA}, a new Rijndael S-box
structure named \emph{Affine-Inverse-Affine} (AIA) is designed
supposed to increase the algebraic complexity of said
S-box. Questioning this claim, we considered it worthwhile to try and
check their calculations and assertions.

A different affine transformation (\ref{e:affx}) with $\Byt{a}=\hexa{5B}$
and $\Byt{b}=\hexa{5D}$ is chosen. This transformation is applied before
and after the inversion step:
\begin{displaymath}
  S_{\mathrm{AIA}}(\Byt{x}) = A \circ I \circ A = A(I(A(\Byt{x})))
\end{displaymath}

Cui \etal~\cite{cui2014:aes} derive a multivariate quadratic equation
system of $S_{\mathrm{AIA}}$ using the coefficients of the polynomial expression
of the S-box.
They write down the equation system with indices for rounds and input
bytes for the AES algorithm (but never use them).
The round indices will be omitted here as they don't matter in what
follows.
As before, by $\Byt{x}$ is denoted the input byte variable of the
S-box function. Intermediate variables are denoted by $\Byt{y}_{0}$,
$\Byt{y}_{1}$, \ldots, $\Byt{y}_{253}$ and the output variable by
$\Byt{z}$. According to the polynomial expression of the new
AIA S-box, the S-box transformation can be described by the
following quadratic equations over GF($2^8$):
\begin{equation}
  \left\{
    \begin{array}{l}
      \Byt{x}\, \Byt{y}_{0} = 1\\[0.3em]
      \Byt{y}_{m}\, \Byt{y}_{0} = \Byt{y}_{m+1}
      \,\textrm{, for $0 \le m \le 252$ and }
      \Byt{y}_{253} = \Byt{x}\\[0.3em]
      \Byt{z} =
        g \left( \Byt{y}_{0}, \Byt{y}_{1}, \ldots, \Byt{y}_{252}, \Byt{x} \right)
    \end{array}
  \right.
  \label{e:AIA1}
\end{equation}
Cui \etal~\cite{cui2014:aes} define the function $g$ by the
polynomial expression of their S-box.
We calculated the coefficients for the polynomial
expression of $S_{\mathrm{AIA}}$ (its Lagrange polynomial) in
Sage\footnote{The Sage code comprises some twenty lines and is
  accessible at SageMathCloud~\cite{SBoxLPinSMC}.}
and tabulate them in Listing~\ref{c:constpols}.
Thereby, the function $g$ reads
\begin{equation}
  \begin{array}{rcl}
    \lefteqn{g\left( \Byt{y}_{0}, \Byt{y}_{1}, \ldots, \Byt{y}_{252},
      \Byt{x} \right) =}
    \\*[0.3ex]
    & & \hexa{FA} + \hexa{12} \Byt{x} + \hexa{26} \Byt{y}_{252}
             + \ldots + \hexa{E5} \Byt{y}_{2}  + \hexa{A9} \Byt{y}_{1}
             + \hexa{A6} \Byt{y}_{0}
  \end{array}
  \label{e:AIA2}
\end{equation}
(The coefficients of Cui \etal~\cite{cui2014:aes} as listed
in their Table~1 which in turn corresponds to Table~3 in Cui, Huang,
\etal~\cite{cui2011:AIA} are wrong and don't represent the polynomial
expression of their S-box $S_{\mathrm{AIA}}$, although, Cui, Huang,
\etal\ list in their preceding Table~2, correctly,
the output values of $S_{\mathrm{AIA}}$.)

The equation system (\ref{e:AIA1}) can be modeled in Sage with the
help of the preparatory code shown in Listing~\ref{c:prepAIA}:
\begin{sageverbatim}%
[caption={Preparation for equation system of AIA S-box},label=c:prepAIA]
  nb = 8
  ny = 253
  varlxz = [c + str(p) for c in 'xz' for p in range(nb)]
  varly = ['y' + str(p) for p in range(nb*ny)]
  B = BooleanPolynomialRing(names = varlxz + varly)
  B.inject_variables()
  P.<p> = PolynomialRing(B)
  Byte.<t> = P.quotient_ring(p^8 + p^4 + p^3 + p + 1)
  X = B.gens()[:nb]
  Z = B.gens()[nb:][:nb]
  YY = [B.gens()[(2+m)*nb:][:nb] for m in range(ny)]
  x = Byte(list(X))
  z = Byte(list(Z))
  yy = [Byte(list(_Y)) for _Y in YY]
\end{sageverbatim}
In lines 3 and 4 lists of strings for the names of coefficients for
the byte variables $\Byt{x}$, $\Byt{z}$, and $\Byt{y}_{0}$, \ldots,
$\Byt{y}_{252}$ are generated.

Then, as in Listing~\ref{c:polyring}, in line 5 a Boolean polynomial
ring for these coefficients is constructed, assigning $GF(2)$
properties to them, in line 6 the coefficients are made available as
variables and in line 7 a polynomial ring over the Boolean polynomial
ring \lstinline|B| is constructed and from that in line 8, eventually,
the quotient ring \lstinline|Byte| with modulus $m(t)$.

In lines 9 to 11 tuples of the coefficient variables of the byte
polynomials are created for convenience. For the $\Byt{y}$-variables
the coefficients are grouped byte-wise in sub-lists.
In the last three lines, finally, the polynomials of the byte variables are
defined using these tuples as arguments for the \lstinline|Byte| constructor.
For the $\Byt{y}$-variables a list of polynomials is used.

In the next Sage code block (Listing~\ref{c:constpols}), the
coefficients of the polynomial expression of $S_{\mathrm{AIA}}$ are
given in hexadecimal notation as a list of strings which is
transformed to a list of the constant polynomials that enter $g$
(\ref{e:AIA2}):
\begin{sageverbatim}%
[numbers=none,caption={Generating constant polynomials of AIA S-box},label=c:constpols]
  sbt = [
   'FA 12 26 E7 9A C7 DB 79 56 01 D3 59 52 ED 97 C9',
   '47 46 FC 7C 5A 50 49 BF F4 F8 63 C8 82 1B EE 74',
   '3B 5D F8 02 2D 64 1A 15 BA DB 59 FE FB D6 97 FF',
   'AB 3F B4 09 32 77 AB 52 4D 96 D5 BB DE 30 DE 05',
   '62 23 7C 69 66 75 9F E9 9B 60 88 2F D1 8F 09 F4',
   '1E EF C4 48 0D A5 AE 7A 38 9B 71 F2 9F 44 B3 99',
   '20 C5 13 12 19 C2 5F 5B AD FA D5 49 7B F8 16 07',
   'B6 75 E9 B0 CA E8 83 C1 4E 75 C5 5E 91 07 86 BF',
   '6F C2 25 35 D3 7F CC 0D AC 7A C9 EC D2 3F C3 21',
   '7E A9 2A 6D A8 66 F8 7D D2 1B FE CD 58 64 25 DA',
   'AE 49 2D 4F 0C 74 F2 42 4A 87 42 9B 83 50 F1 91',
   'C1 02 4F 2A C9 19 37 59 D5 74 8D 0B 20 C5 AF 28',
   '47 FB 09 87 10 6A 3B C8 8B 08 5B 8B 13 0E 73 7E',
   'FA 45 85 18 D5 90 4E 71 E6 F2 BF EE 30 E9 99 54',
   '30 63 8F 03 92 91 0C 43 09 66 E5 76 6A 93 87 E4',
   '6C 6A 87 A1 CB 64 AA 5C FB 05 5A DE E5 A9 A6 00']
  sbt = ' '.join(sbt).split()
  sbp = [Byte(ZZ(_, 16).bits()) for _ in sbt]
\end{sageverbatim}
In the last line of this code block each two-digit hexadecimal number
in the table represented by a two character string is converted into
a decimal number by the code fragment \lstinline|ZZ(_, 16)|.
Appending \lstinline|.bits()| transforms it into a list of
0s and 1s, a big-endian binary representation of the
hexadecimal number. Applying the \lstinline|Byte| constructor
gives the corresponding constant polynomial.

With these preparations the equation system (\ref{e:AIA1}),
(\ref{e:AIA2}) of the AIA S-box
(equation~9 in  Cui \etal~\cite{cui2014:aes}) can be
modeled in Sage as shown in Listing~\ref{c:eqsAIA}.
\begin{sageverbatim}%
[numbers=none,caption={Equation system of AIA S-box in Sage},label=c:eqsAIA]
  g = sbp[0] + sbp[1] * x \
      + sum(sbp[2+m] * yy[ny-1-m] for m in range(ny))
  yy.append(x)
  E9 = [x * yy[0] + 1]
  E9.extend(yy[m] * yy[0] + yy[m+1] for m in range(ny))
  E9.append(z + g)
  mq3 = flatten([_.list() for _ in E9])[1:]
\end{sageverbatim}
In the last line the first term (the $t^0$-coefficient of
$\Byt{x}\Byt{y_{0}}+1$) is discarded since it is only true with
probability 255/256 (false if $\Byt{x} = 0$).
Using this Sage model we evaluate some metrics of this MQ as before.
This gives for the new AIA S-box 2,039 equations, with
between 3 and 1,034 monomials per equation. These equations have in
total 18,232 different monomials.
In order to apply the criteria
of section~\ref{s:Algebraicattacks},
the intermediate variables also were taken into account by interpreting the
parameter $n$ in the definitions (\ref{e:gamma}) and (\ref{e:gammaCP})
as the number of dependent variables. Hence, the number $8\times 254$
is used as $n$, and not only 8.
This results in $\Gamma = (16,193/2,032)^{8} \approx\ 2^{24.0}$ and
$\Gamma_{\mathrm{CP}} = (18,232/2,032)^{9} \approx 2^{28.5}$ as
estimation for RAA.\footnote{Formal
  application of $n = 8$ yields unlikely high values:
  $\Gamma \approx 2^{22,241}$ and $ \Gamma_{\mathrm{CP}} \approx 2^{100}$.}

Also, the CPU time to evaluate all 256 solutions of this MQ with a SAT
solver is ca.\ 7~s, which is 12 ($\approx 2^4$) times as long as
for the original Rijndael S-box.

Cui \etal~\cite{cui2014:aes} count a totally different number of
equations and terms based on the byte variables, not on their
polynomial coefficients, which contradicts the scheme applied to the
Rijndael S-box with which they compare, and therefore, is misleading.

Further, this method used by Cui \etal~\cite{cui2014:aes} to derive
an MQ for their AIA S-box applies to any S-box using the coefficients
of its polynomial expression.
This illustrates that the resulting numbers of equations and terms are deceptive
as criterion for the estimation of algebraic attack resistance and
inapt to differentiate the quality of byte S-boxes.
To substantiate this point, we have derived such an MQ for the original
Rijndael S-box by using its polynomial expression in equation
(\ref{e:AIA1}). The function $g$ then reads
\begin{equation}
  \begin{array}{rcl}
    \lefteqn{g_{\mathrm{SRD}}\left( \Byt{y}_{0}, \Byt{y}_{1}, \ldots,
      \Byt{y}_{252}, \Byt{x} \right) =}
    \\*[0.3ex]
    & & \hexa{63} + \hexa{8F} \Byt{y}_{127}
        + \hexa{B5} \Byt{y}_{63} + \hexa{01} \Byt{y}_{31} + \hexa{F4}
        \Byt{y}_{15} + \hexa{25} \Byt{y}_{7} \, +\\
    & &
        \hexa{F9} \Byt{y}_{3} + \hexa{09} \Byt{y}_{1} + \hexa{05} \Byt{y}_{0}
  \end{array}
  \label{e:gSRD}
\end{equation}
The same Sage code (Listings~\ref{c:prepAIA}, \ref{c:constpols}, and
\ref{c:eqsAIA}) was used with an adapted table of the polynomial
expression coefficients in accordance with equation (\ref{e:gSRD}).
%
This MQ of the Rijndael S-box exhibits the same number
of equations with the same number of different monomials as the MQ of
$S_{\mathrm{AIA}}$ resulting in equally high, misleading $\Gamma$ values.
Also, the SAT solver needs the same 7~s CPU time to find the solutions
of this MQ.

In contrast, we will show in the next section how to derive, aided by
computer mathematics, a much simpler MQ for
the AIA S-box which shows that its resistance against algebraic
attacks according to the effort of a SAT solver not really exceeds
that of the original Rijndael S-box. But the RAA criteria
exaggerate the hardness of that simpler MQ. 
\section{Concise MQ for the AIA S-box in Sage}
\label{s:MQforSAIA}
Building on the Sage code presented so far we derive a much simpler MQ
for the AIA S-box. Its resistance against algebraic attacks according
to the RAA criterion should be greater than that of the original
Rijndael S-box but over-estimates the time it takes to solve the
system with a SAT solver.

For the inversion step we now use, temporarily, two intermediate byte
variables $\Byt{y}_{0}$ and $\Byt{y}_{1}$, named \lstinline|yy[0]| and
\lstinline|yy[1]| in the Sage code. Their coefficients shall be $y_{0},
\ldots, y_{7}$ and $y_{8}, \ldots, y_{15}$, respectively.
The three steps of the AIA S-box are
\begin{displaymath}
  \Byt{z} = A(\Byt{y}_{1})\, ,\
  \Byt{y}_{1} = I(\Byt{y}_{0})\, ,\
  \Byt{y}_{0} = A(\Byt{x})
\end{displaymath}
Beginning with the inversion step we first model
\begin{equation}
  \left\{ \
  \begin{array}{lcl}
    \Byt{y}_{0}^{128} & = & \Byt{y}_{1}^{128} \Byt{y}_{0}\\
    \Byt{y}_{1}^{128} & = & \Byt{y}_{0}^{128} \Byt{y}_{1}\\
    \Byt{y}_{0}^{3} & = & \Byt{y}_{0}^{4} \Byt{y}_{1}\\
    \Byt{y}_{1}^{3} & = & \Byt{y}_{1}^{4} \Byt{y}_{0}
  \end{array}
  \right.
  \label{e:Inv32}
\end{equation}
The last two equations in (\ref{e:Inv32}) are the only
other additional (fully quadratic) MQ (besides $\Byt{y}_{0}\Byt{y}_{1} = 1$)
for the inversion as stated already by Courtois and Pieprzyk in the
extended version of~\cite{courtois2002:algres} (compare also Cheon and
Lee~\cite{cheon2004:algres}).
These additional equations are necessary to completely define the
S-box $S_{\mathrm{AIA}}$.
Without them the system is under-defined, as, for example, the solution
with a SAT solver shows.
In Sage the equations (\ref{e:Inv32}) read
%
\begin{sageverbatim}
  E10 = yy[0]^128 + yy[1]^128 * yy[0]
  E11 = yy[1]^128 + yy[0]^128 * yy[1]
  E12 = yy[0]^3 + yy[0]^4 * yy[1]
  E13 = yy[1]^3 + yy[1]^4 * yy[0]
\end{sageverbatim}
The linear transformations according to equation (\ref{e:affx}) are
\begin{eqnarray}
  \Byt{y}_{0} & = & \Byt{a}\, \Byt{x} \mod (t^8 + 1) + \Byt{b}
  \label{e:affx4a}
  \\
  \Byt{y}_{1} & = & \Byt{a}^{7} (\Byt{z} + \Byt{b}) \mod (t^8 + 1)
  \label{e:invaffx4a}
\end{eqnarray}
with $\Byt{a}=\hexa{5B}$
(hence $\Byt{a}^{8} \mod (t^8 + 1) = 1$)
and $\Byt{b}=\hexa{5D}$.
In Sage we formulate
\begin{sageverbatim}[caption={Generate dictionary to substitute $\Byt{y}$ variables},label=c:affxAIA]
  Baff.<u> = P.quotient_ring(p^8 + 1)
  a = Baff(ZZ('0x5B').bits())
  b = Baff(ZZ('0x5D').bits())
  eqs14 = dict(zip(YY[0], a * Baff(x) + b))
  eqs14.update(zip(YY[1], a^7 * (Baff(z) + b)))
\end{sageverbatim}
The right hand sides of equations (\ref{e:affx4a}) and
(\ref{e:invaffx4a}) enter Listing~\ref{c:affxAIA} in the two last lines.
Both their coefficients are inserted into the same
dictionary \lstinline|eqs14|.
%
Applying this substitutions in Sage to get rid of the
intermediate byte variables is straight forward
\begin{sageverbatim}
  eqs10 = [_.subs(eqs14) for _ in E10.list()]
  eqs11 = [_.subs(eqs14) for _ in E11.list()]
  eqs12 = [_.subs(eqs14) for _ in E12.list()]
  eqs13 = [_.subs(eqs14) for _ in E13.list()]
\end{sageverbatim}
and gives already the final, concise MQ for the S-box $S_{\mathrm{AIA}}$:
\begin{sageverbatim}
  mq5 = eqs10 + eqs11 + eqs12 + eqs13
\end{sageverbatim}
For this MQ counting the numbers of equations and monomials in Sage is
done as before.  It has 32 equations, with between 33 and 60 monomials
per equation and 137 different monomials in total.
This makes according to definitions (\ref{e:gamma}) and
(\ref{e:gammaCP})
$\Gamma = (105/8)^{14} \approx 2^{52}$ and
$\Gamma_{\mathrm{CP}} = (137/8)^{5} \approx 2^{20.5}$.

Nonetheless, the SAT solver takes ca.\ 0.8~s CPU time on the same
computer (2.8 GHz CPU, 8 GB RAM) to find all and only 256 solutions
for this MQ.
The evaluation of its Gröbner basis takes ca. 16 s.
The GB has 8 equations of degree 7 with 263 different monomials and
its solution with a SAT solver is obtained as fast as that of the GB
of the Rijndael S-box. Clearly, the values of the hardness criteria
do not correlate with the effort of the SAT solver for this MQ.

For comparison and as an additional reference value for the RAA
estimation, we have derived such an MQ with 32 equations for the
original Rijndael S-box using the four equations (\ref{e:Inv32})
(replacing $\Byt{y}_{0}$ by $\Byt{x}$ and $\Byt{y}_{1}$ by $\Byt{y}$)
and its affine transformation (\ref{e:invaffx}) (with
$\Byt{a}=\hexa{1F}$, $\Byt{b}=\hexa{63}$). This MQ has the same number
of equations and the same number of different monomials, thus, the
same values for the hardness criteria as that of
$S_{\mathrm{AIA}}$. The solution with the SAT solver of this MQ for
the original Rijndael S-box takes the same CPU time as the
solution of the 32 equation MQ of $S_{\mathrm{AIA}}$, namely,
ca.\ 0.8~s.
This shows how Sage can easily be used to disprove the practicality
of the hardness criteria.
\section{Conclusion}
SageMath is a very appropriate, powerful computer mathematics tool to
analyze cryptographic problems formulated with byte variables as
polynomials in a quotient ring. Sage draws its strength in this area
mainly from the integration of the BRiAl, former PolyBoRi,
library~\cite{stein2015:sage,brickenstein2009:PolyBoRi}.

We have used Sage to demonstrate how to produce various polynomial
Multivariate Quadratic equation systems (MQ) as well as their Gröbner
basis for the Rijndael S-box S$_{\mathrm{RD}}$ and similar S-boxes in
a simple and straightforward manner.
Using the flexible structures and interface of Sage one can easily
evaluate metrics of the resulting polynomial systems, like the
number of different monomials in the system, the length of the
equations, the frequency of the appearance of certain terms or
variables in equations etc.
With this facility we generated the necessary inputs for the
application of estimations of the Resistance against Algebraic Attacks
(RAA) proposed by Cui \etal~\cite{cui2014:aes} or by Courtois
and Pieprzyk~\cite{courtois2002:algres}. Parallelly, we performed
numerical experiments by solving the corresponding MQ with a SAT
solver using the required computing time as a measure for the
RAA.

Our results in this respect are revealing. We couldn't validate the
predicted reduced hardness for a system to solve with increasing
number of equations and number of independent monomials in the
polynomial systems which both the formulas forecast. There is in fact
a slight increased solver effort when one reduces from the 23
Rijndael S-box equations to the 16 but quantitatively this is badly
represented in both formulas.

Cui~\etal\ constructed a complicated algebraic expression for a new
Rijndael S-box S$_{\mathrm{AIA}}$ (starting from its Lagrange
polynomial expression with 255 coefficients) which necessarily leads
to a great number of equations and independent monomials in the
resulting MQ. On this basis they thought they have demonstrated a
remarkable new S-box practically not possible to solve according to
the here discussed and by us dismissed RAA formulas. However there are
gaps in their concept arising from inconsistency in their comparison
principle as well as the lack of thoroughness in the investigation of
the properties of the new algebraic expression which we showed can be
equivalently written in a much simpler form leading to a polynomial
system nearly as easy solved as that of the original Rijndael S-box.

We also mapped the original Rijndael S-box with its 9 Lagrange
coefficients on the AIA form of Cui~\etal\ which gave us as result
the same \emph{huge} number of variables and multitude of polynomials which
should manifest that this is no way to create especially hard
cryptographic S-boxes.

We presented how to show with SageMath that the, by Cui~\etal\ so called,
\emph{improved} S-box S$_{\mathrm{AIA}}$ is in fact not even
marginally an improvement.

\tablename~\ref{t:CompCrit} gives a survey
\begin{table}[ht]
\begin{center}
\begin{tabular}{l|cccc|c}
  \hline\hline
  \multirow{2}{*}{S-box} & \multicolumn{4}{c|}{MQ}   & Gröbner\\
                         & RD & RD & AIA/RD & AIA/RD & AIA/RD\\
  \hline
  maximal degree            &     2 &     2 &      2 &    2 &    7\\
  \# equations              &    23 &    16 &  2,039 &   32 &    8\\
  \# monomials              &    81 &    81 & 18,232 &  137 &  263\\
  \# dependent variables    &     8 &     8 &  2,032 &    8 &    8\\
  $\log_{2}(\Gamma)$        &  22.9 &  27.2 &   24.0 & 52.0 & --\\
  $\log_{2}(\Gamma_{\mathrm{CP}})$ & 13.4 &  20.0 & 28.5 & 20.5 & --\\
  SAT solver CPU time       & 0.6 s & 0.7 s &    7 s & 0.8 s & 0.15 s\\ 
  \hline\hline
\end{tabular}
\end{center}
\caption{\label{t:CompCrit}%
  Survey of S-box MQ and estimations of their RAA.}
\end{table}
%
%
of the MQ and the results of the algebraic attack resistance estimations
scrutinized in this work.

We conclude, that in order to assess the resistance of an S-box
against algebraic attacks it is not sufficient to derive some
multivariate quadratic equation system and analyze it.
Instead one would have to show that the derived MQ is optimal and
superior to its Gröbner basis for solving and, thus, attacking it or
the cipher it is used in.
%
%

%
\end{document}